\documentclass{pasj00}
\onecolumn

\begin{document}
\SetRunningHead{}{Large Scale Clustering of SDSS QSOs}
\Received{2004/12/03}
\Accepted{2005/05/09}

\title{Large Scale Clustering of Sloan Digital
Sky Survey Quasars: Impact of the Baryon Density
and the Cosmological Constant}
\author{%
Kazuhiro \textsc{Yahata}\altaffilmark{1},
Yasushi \textsc{Suto}\altaffilmark{1},
Issha \textsc{Kayo}\altaffilmark{1,2},
Takahiko \textsc{Matsubara}\altaffilmark{2},\\
Andrew \textsc{Connolly}\altaffilmark{3},
Daniel \textsc{Vanden Berk}\altaffilmark{3,4},
Ravi \textsc{Sheth}\altaffilmark{3,5}, \\
Istv\'an \textsc{Szapudi}\altaffilmark{6},
Scott F. \textsc{Anderson}\altaffilmark{7},
Neta \textsc{Bahcall}\altaffilmark{8},
Jon \textsc{Brinkmann}\altaffilmark{9}, \\
Istv\'an \textsc{Csabai}\altaffilmark{10},
Xiaohui \textsc{Fan}\altaffilmark{11},
Jon \textsc{Loveday}\altaffilmark{12},\\
Alexander S. \textsc{Szalay}\altaffilmark{10}, and,
Donald \textsc{York}\altaffilmark{13}}
\altaffiltext{1}
{Department of Physics, School of Science,
The University of Tokyo, Tokyo 113-0033}
\altaffiltext{2}
{Department of Physics and Astrophysics,
Nagoya University, Chikusa, Nagoya 161-8602}
\altaffiltext{3}
{University of Pittsburgh, Department of Physics and
Astronomy, \\ 3941 O'Hara Street, Pittsburgh, PA 15260, USA}
\altaffiltext{4}
{Department of Astronomy and Astrophysics, Pennsylvania State
University, \\ 525 Davey Laboratory, University Park, PA 16802, USA}
\altaffiltext{5}
{Department of Physics and Astronomy, University of Pennsylvania, \\
Philadelphia, PA 15104, USA}
\altaffiltext{6}
{Institute for Astronomy, University of Hawaii, \\ 2680
Woodlawn Drive, Honolulu, HI 96822, USA}
\altaffiltext{7}{Astronomy Department, University of Washington,
Box 351580, Seattle, WA 98195, USA}
\altaffiltext{8}
{Princeton University Observatory, Peyton Hall, Princeton, NJ 08544,  
USA}
\altaffiltext{9}
{Apache Point Observatory, Sunspot, NM 88349-0059, USA}
\altaffiltext{10}
{Department of Physics and Astronomy, The Johns Hopkins University, \\
3701 San Martin Drive, Baltimore, MD 21218, USA}
\altaffiltext{11}{Steward Observatory, The University of Arizona,
Tucson, AZ 85721, USA}
\altaffiltext{12}
{Astronomy Centre, University of Sussex, Falmer, Brighton BN1 9QJ, UK}
\altaffiltext{13}{Astronomy and Astrophysics Department, University of
Chicago, \\ 5640 South Ellis Avenue, Chicago, IL 60637, USA}
\email{yahata@utap.phys.s.u-tokyo.ac.jp}
\maketitle

\begin{abstract}
We report the first result of the clustering analysis of Sloan Digital
Sky Survey (SDSS) quasars.  We compute the two-point correlation
function (2PCF) of SDSS quasars in redshift space at $8h^{-1}{\rm Mpc}
< s < 500h^{-1}{\rm Mpc}$, with particular attention to its baryonic
signature.  Our sample consists of 19986 quasars extracted from the
SDSS Data Release 4 (DR4).  The redshift range of the sample is $0.72
\le z \le 2.24$ (the mean redshift is $\bar z = 1.46$) and the reddening-corrected $i$-band apparent
magnitude range is $15.0 \le m_{i,{\rm rc}} \le 19.1$.  Due to the
relatively low number density of the quasar sample, the bump in the
power spectrum due to the baryon density, $\Omega_{\rm b}$, is not
clearly visible. The effect of the baryon density is, however, to
distort the overall shape of the 2PCF. The
degree of distortion makes it an
interesting alternate measure of the baryonic signature.  Assuming a
scale-independent linear bias and the spatially flat universe, i.e.,
$\Omega_{\rm b} + \Omega_{\rm d} + \Omega_\Lambda =1$, where
$\Omega_{\rm d}$ and $\Omega_\Lambda$ denote the density parameters of
dark matter and the cosmological constant, we combine the observed
quasar 2PCF and the predicted matter 2PCF to put constraints on
$\Omega_{\rm b}$ and $\Omega_\Lambda$.

Our result is fitted as $0.80- 2.8\Omega_{\rm b} < \Omega_\Lambda <
0.90 - 1.4\Omega_{\rm b}$ at the 2$\sigma$ confidence level, which is
consistent with results from other cosmological observations such as
WMAP.  The ``mean'' bias parameter of our quasar sample is
$1.59{\sigma_8}^{-1}$ (for $\Omega_b=0.04$ and $\Omega_\Lambda=0.7$),
where $\sigma_8$ is the top-hat mass fluctuation amplitude at
$8h^{-1}$Mpc.  We also estimate the corresponding bias parameter of
quasars at $z=0$, $b_{\rm QSO,Fry}(0)$, assuming Fry's bias evolution
model. For $\Omega_{\rm b}=0.04$, $\Omega_\Lambda=0.73$ and
$\Omega_{\rm d}=0.23$, we find $b_{\rm
QSO,Fry}(0)=0.54+0.83{\sigma_8}^{-1}$ which is valid for $0.6
<\sigma_8 < 1.0$.
\end{abstract}

\section{Introduction}
Quasars are intrinsically bright and preferentially located at
relatively high redshift. They are, therefore, naturally suited for
exploring the structure of the distant universe out to redshifts $z
\sim 3$, a regime not easily accessible from existing wide-field
galaxy surveys.  While surveys of the universe up to $z\sim 6$ have
significantly advanced in recent years using Lyman break-galaxies and
Lyman-$\alpha$ emitters (e.g., \cite{Steidel99, Shimasaku03}; Ouchi et
al. 2004a; Ouchi et al. 2004b),
they remain limited in terms of the areas they cover (barely extending
to $O(100)h^{-1}{\rm Mpc}$). As such they are not well suited to
measuring the large scale clustering at high redshift (i.e.\ they are
sample variance limited). Quasar samples, in contrast, can be surveyed
over very large volumes and, therefore, play an important and
complementary role in measuring the evolution of structure in the
universe on the largest accessible scales.

The primary goal of this quasar clustering analysis is to directly
probe and quantify the large-scale structure of the universe on scales
up to a few $h^{-1}$Gpc, where $h$ denotes the Hubble constant in
units of $100 {\rm km s^{-1}Mpc^{-1}}$.  Eventually such analyses may
enable us to answer fundamental questions including the homogeneity of
the universe on Gpc scales or the validity of the cosmological
principle (e.g., \cite{Lahav04}).  Also the clustering signal provides
important clues to the formation and evolution of quasars themselves
and may shed light on the growth history of their central massive
black holes.

Although there are several seminal papers addressing quasar clustering
(\cite{Osmer1981, Shanks1987, Boyle1993, Croom1996}), those previous
quasar samples were fairly small which limits their statistical power
in constraining large scale structure. The 2dF QSO Redshift Survey
(2QZ:\cite{2QZ-2,2QZ-9, 2QZ-11}) and the SDSS are among the largest
statistical samples which should significantly advance the statistics
of quasar clustering analysis.

This is the first of a series of SDSS quasar clustering papers and is
aimed at exploring the impact of the baryon density and the cosmological
constant on the quasar 2PCF on scales of $\sim 100h^{-1}$Mpc.  In this
sense, our current approach is quite complementary to the existing
galaxy clustering studies by the SDSS collaboration (e.g., Tegmark et
al. 2004; Pope et al. 2004).  The samples and techniques are
complementary: the galaxies and quasars are selected using different
algorithms, they explore vastly different volumes, the number densities
of the objects are very different, and they cover very different
redshift ranges.  It would be interesting to see if both data sets yield
consistent results on constraints on the cosmological parameters.

We will address the dependence of the quasar 2PCFs on redshift and
luminosity and comparison with those of galaxies and AGNs in a separate
paper \citep{connolly05}.  To be specific, we compute the 2PCF of
quasars from SDSS Data Release 4 (DR4) on scales from 8$h^{-1}$Mpc to
2$h^{-1}$Gpc and perform statistical analyses that pay particular
attention to the embedded baryonic signature. From recent progress in
the theoretical and observational understanding of the initial
conditions present in the early universe, the 2PCF of the matter is well
described. Two characteristic scales in the matter 2PCF, $\xi_m(r,z)$,
are present at approximately 100$h^{-1}{\rm Mpc}$; the baryon bump and
the zero-crossing point (denoted by $s_{\rm zero}$ in what follows).
These features are sensitive to the baryon density parameter,
$\Omega_{\rm b}$, and the matter density parameter, $\Omega_{\rm m}$,
where we define $\Omega_{\rm m} \equiv \Omega_{\rm b} + \Omega_{\rm d}$,
with $\Omega_{\rm d}$ being the density parameter of cold dark matter
(CDM).

We derive constraints on $\Omega_{\rm b}$ and $\Omega_\Lambda$ assuming
a spatially flat universe (i.e., $\Omega_{\rm b} + \Omega_{\rm d} +
\Omega_\Lambda = 1$) throughout this paper.  Our attempt may be
partially interpreted as yet another cosmological test to determine the
values of $\Omega_{\rm b}$ and $\Omega_\Lambda$ independent of existing
methods. In this respect, our current results discussed below are not
stringent enough to improve the previous constraints. Nevertheless we
believe that the attempt to look for the baryonic signature in the
cosmological distribution of quasars is of scientific interest on its
own.

\section{SDSS quasar sample}
\subsection{The Sloan Digital Sky Survey}
The Sloan Digital Sky Survey (SDSS;\cite{Y2000} for a technical
summary) provides the largest homogeneous sample of quasars currently
available, and thus is best suited for our current purpose. The
wide-field five band photometric system
(\cite{Gunn,Fukugita1996,Hogg2001,Ivezic2004,Smith2002}) enables the
identification of quasars with high efficiency and completeness.  The
precision astrometry \citep{Pier2003} maintains the accuracy of the
clustering analysis even in 100$h^{-1}{\rm kpc}$ scale.  The SDSS is
also excellent in terms of the homogeneity of the spectroscopic sample
because its spectroscopic targets are selected consistently on the
basis of their photometric data.  The details of quasar target
selection algorithm are described by \citet{targetselection}, and the
spectroscopic tiling algorithm is described by \citet{tiling}.

\subsection{Quasar sample selection}

The spectroscopic sample of quasars in the DR4 is selected according
to several different spectroscopic target selection criteria (see
Table 27 of \cite{EDR}).  We consider only the point sources, and
extract our quasar sample for the current analysis on the basis of
four conditions; first, to ensure a homogeneity of the sample, we
choose those spectroscopic quasars whose ``primTarget'' flag (see
Table 6 of \cite{EDR}) indicates either ``QSO\_CAP'' or ``QSO\_SKIRT''
(see Fig.1 and section 3.4 of \cite{targetselection}).  Second, their
$i$-band magnitude corrected for the Galactic reddening should satisfy
$15.0 < m_{i,{\rm rc}} < 19.1$, which are the limits used by the
quasar target selection algorithm.  We adopt the reddening correction
of \citet{reddening}.  Although this correction is calculated adopting
the typical spectrum energy distribution of elliptical galaxies and
the old SDSS filter response functions the resulting difference is
negligibly small.  Third, the redshift limits of this sample are $0.72
< z < 2.24$.  In the SDSS photometric system, the color locus of
quasars in the redshift range $2.4 < z < 3.0$ crosses the stellar
locus and thus the survey completeness becomes significantly degraded
in that redshift range. This defines the upper redshift limit of our
sample.  The lower limit is chosen so that the absolute magnitude of
all the selected quasars should be brighter than -23 in i-band.
Finally, we exclude the southern sky region because the selection of
the region is different from that of the northern sky. The total
number of selected quasars is 19986.

The SDSS photometric data have two different versions; ``BEST'' version
denotes the highest quality data at the time of the data release, while
``TARGET'' version is the data at the time the target selection algorithm
was run for that part of the sky (i.e., the photometric data may be
updated at the time of the data release).  Our current sample uses the
``primTarget'' flag which is based on the ``TARGET'' version.  In
reality, however, the two versions may lead to some systematic effect
due to the mismatch of the time-dependent selection function.  We
compared BEST and TARGET using the SDSS Data Release 3 (DR3: see
\cite{dr3}) QSO catalog \citep{Schneider05}, and found that the
difference does not affect the resulting 2PCFs within the current
errorbars (see Fig.\ref{fig:sdss_n1} below).  We note that our sample in
the DR3 region slightly differs from DR3 QSO catalog \citep{Schneider05}
because we have not manually inspected all objects classified by the
{\it pipeline} as quasars.  The number of quasars in a subset of our DR4
sample that is located in the DR3 region is 16656, out of which 32
objects are not listed in the catalog of \citet{Schneider05} and 9
objects have incorrect redshifts. This difference of the quasar samples,
however, has a negligible effect on the estimated 2PCF as will be shown
below.

\begin{figure}[p]
   \centering \FigureFile(140mm,70mm){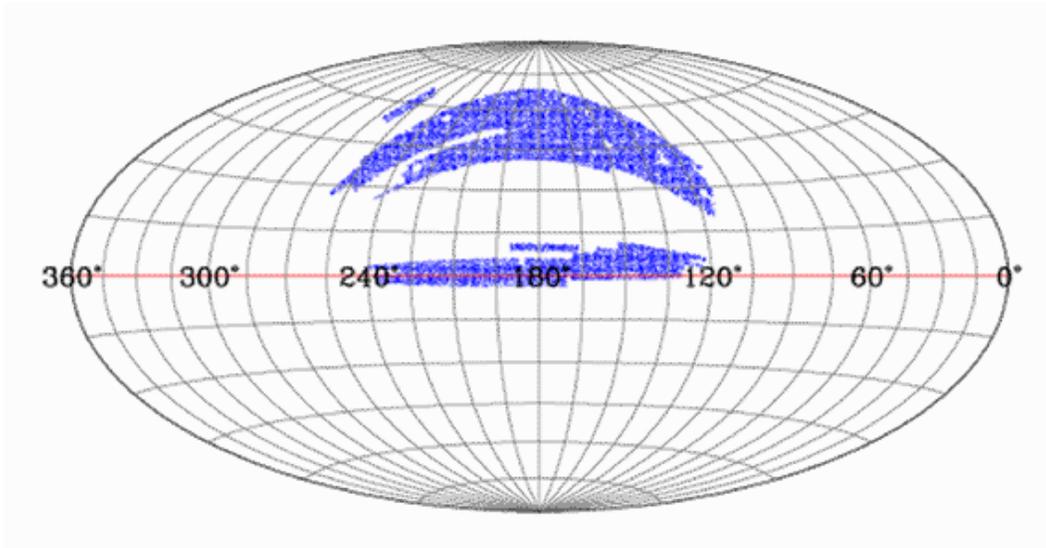}
\caption{Equatorial coordinate distribution of SDSS quasars that we use
in the current analysis (19986 in total).
\label{fig:quasar-distribution}}
\end{figure}

The equatorial coordinate distribution of our quasar sample is plotted in
Figure \ref{fig:quasar-distribution}, and a projected wedge diagram of the
quasars is shown in Figure \ref{fig:wedge_quasar}.  Figure \ref{fig:M_z}
shows the scatter plot of redshift and $i$-band absolute magnitude
({\it Upper panel}) and the redshift histogram ({\it Lower panel}).

Giving the errorbars in Fig.\ref{fig:sdss_n1}, the photometric completeness, fiber collisions and the effect of errors in the reddening correction are not supposed to change our result.  
Those issues will be fully addressed by \citet{connolly05} who found
that none appears to significantly affect the quasar 2PCFs presented
below.
\subsection{Analysis}
  We use the Landy-Szalay estimator \citep{estimator} in computing the
2PCF from the quasar sample:
\begin{eqnarray}
\label{eq:LSestimator}
\xi_{\rm QSO,obs}(s_i\le s < s_{i+1})
= \frac{QQ(s_i\le s < s_{i+1})}{RR(s_i\le s < s_{i+1})}
-2 \frac{QR(s_i\le s < s_{i+1})}{RR(s_i\le s < s_{i+1})}
+1,
\end{eqnarray}
where $QQ$, $QR$, and $RR$ are the number of corresponding pairs of
quasars and random particles that are normalized by $N_Q(N_Q-1)/2$, $N_Q
N_R$, $N_R(N_R-1)/2$, respectively ($N_Q$ and $N_R$ are the numbers of
quasars and random particles). The accuracy and convergent properties of
the Landy -- Szalay estimator are discussed in the appendix of
\citet{Kayo-3pt} and in \citet{Kerscher00}.

The redshift distribution of random particles is determined so as to
reproduce that of the quasars.  The latter is estimated by averaging
over a finite width of redshift, $\Delta z = 0.08$.  We find that
other choices of $\Delta z = 0.02$, $\Delta z = 0.04$ and $\Delta z =
0.10$ result in almost identical $\xi_{\rm QSO,obs}(s)$.  The number
of random particles, $N_R$, is about 600,000, which is found to
provide robust estimates of $\xi_{\rm QSO,obs}(s)$ on the scales $s>8
h^{-1}{\rm Mpc}$.  We evaluate the covariance matrix, $C_{ij}$, from
40 jack-knife re-samplings (see e.g., \cite{Lupton1993}).  In order to
ensure the statistical significance, the comparison between the
observed and the predicted 2PCFs needs to be done after the finite
binning of the pair-separation. Empirically, we compute the 2PCFs at 6
different binning offsets; the $i$-th boundary of the $n$-th binning
offset ($n=1 \ldots 6$) is defined as
\begin{equation}
\label{eq:s_in}
s_i(n) = 10^{0.3i + 0.05(n-1)} h^{-1} {\rm Mpc}.
\end{equation}
We then compute the averaged 2PCFs of theoretical predictions
(described in section 3) as follows:
\begin{equation}
\xi_{\rm QSO}(s_i\le s < s_{i+1}) =
  \frac{\int^{s_{i+1}}_{s_i}\xi_{\rm QSO}(s')s'^2 ds'}
{\int^{s_{i+1}}_{s_i}s'^2ds'}.
\end{equation}
In practice, for a given $n$-th offset of the binning, we use bins whose
boundaries satisfy $7.9 h^{-1} {\rm Mpc} < s_i < 502 h^{-1} {\rm Mpc}$
for the following analysis.  Figure \ref{fig:sdss_n1} plots the 2PCFs of
our samples for the $n=1$ binning offset adopting $\Omega_{\rm m} = 0.3$,
$\Omega_\Lambda =0.7$ and $h=0.7$.  We also plot the 2PCFs of a subset
of our DR4 sample that is located in the DR3 region, and two 2PCFs of BEST
and TARGET samples of the DR3 quasar catalog \citep{Schneider05} for
comparison.  The four sets of 2PCFs agree with one another to within
1$\sigma$.  Throughout the paper, we show the 2PCFs in log-log
plots. In order to show the negative part of 2PCFs, we plot the range of
$-10^{-2} < \xi< -10^{-5}$ in the lower panel of each figure by taking
the logarithm of its absolute value.

\section{Theoretical predictions}

\subsection{Parameter dependence of the matter 2PCF}

Since the baryon bump is the counterpart of the acoustic oscillation
peaks in a matter power spectrum, the analysis using the 2PCF should, in
principle, be equivalent to that with the power spectrum.  Nevertheless
we perform the analysis with the 2PCF for the following two reasons.
The first advantage is based on the different behavior of the baryonic
signature in the power spectrum and 2PCF.  While the acoustic
oscillation peaks in a matter power spectrum show up in several
overtones, they translate coherently to the baryon bump in the 2PCF
roughly at one specific scale.  Thus the resulting signature is expected
to be more prominent in the 2PCF.  The second is related to the
inevitably complicated geometry of the survey volume boundary.  Strictly
speaking, the estimate of the power spectrum from the observed sample
has to assume a periodic boundary condition unless resorting to a very
sophisticated estimation methodology (e.g., \cite{Pope04}).  On the other hand,
the 2PCF can be reliably estimated by taking account of the arbitrary
boundary shape of the survey volume in a relatively straightforward
fashion.  For this purpose, we simply distribute random particles over
the survey volume according to the selection function of quasars (the
lower-panel of Fig.\ref{fig:M_z}) and perform standard pair-counting.
This is why we adopt the 2PCF in the present analysis.

In locating the baryon bump and the zero-crossing point, it is important
to have accurate predictions of the matter 2PCF. Here we adopt an
empirical fitting formula for the CDM transfer function incorporating
the effect of $\Omega_{\rm b}$ \citep{EisensteinHu}. The theoretical
model of 2PCFs depends on $\Omega_{\rm m}$ and $\Omega_{\rm b}$ and not
on $\Omega_\Lambda$, while the estimate of the observed 2PCFs requires
the values of $\Omega_{\rm m}$ and $\Omega_\Lambda$. To simplify the
analysis, therefore, we adopt a spatially-flat universe, $\Omega_{\rm m}
+ \Omega_\Lambda =1$, and choose $\Omega_{\rm b}$ and $\Omega_\Lambda$
as two model parameters.

Figure \ref{fig:baryon} shows $\xi_{\rm m}(r,z=0)$ for different values
of $\Omega_{\rm b}$ ($\Omega_\Lambda$ is fixed to $0.73$).
   Baryon bumps are always visible around
$100h^{-1}$Mpc (except for the case of $\Omega_{\rm b}=0$) and their
height and position are sensitive to the value of $\Omega_{\rm b}$.
Figure \ref{fig:lambda} also plots $\xi_{\rm m}(r,z=0)$ but for
different values of $\Omega_\Lambda$ assuming $\Omega_{\rm b}=0.04$.  It
is clear that $s_{\rm zero}$ is very sensitive to $\Omega_\Lambda$ and,
to a lesser extent, to $\Omega_{\rm b}$.

Figure \ref{fig:hubble} shows the dependence of $\xi_{\rm m}(r,z=0)$ on
$h$.  Interestingly, $s_{\rm zero}$ is quite insensitive to the value of
the Hubble constant. If $\Omega_{\rm b}=0$, the CDM linear matter  
spectrum is
approximately given as a universal function of the scaled wavenumber, $q
\equiv k/(\Omega_{\rm m}h^2)$ implying the scaling of $s_{\rm zero}
[h^{-1} {\rm Mpc}] \propto (\Omega_{\rm m}h)^{-1}$ (see also
\cite{matsubara04} for related discussion on the parameter dependence).
We made sure that this scaling holds for $\Omega_{\rm b}<0.01$ but is  
replaced
by a different empirical relation of $s_{\rm zero} [h^{-1} {\rm Mpc}]
\propto \Omega_{\rm m}^{-1} = (1-\Omega_\Lambda)^{-1}$ for
$\Omega_{\rm b}>0.02$ (the second equality comes from our assumption of  
the
spatially flat universe).  We note here that the above feature may be
very useful in breaking the degeneracy in the constraints on  
$\Omega_{\rm m}$
and $h$.
\subsection{Predicting 2PCF of quasars on the light-cone}

In order to predict the observable 2PCF of quasars $\xi_{\rm QSO}(s)$
from the matter 2PCF $\xi_{\rm m}(r,z)$, one has to take into account
three major effects, (i) the light-cone effect, (ii)
redshift-space distortion and (iii) quasar biasing.  Since the
formulation and the prescription of incorporating these effects have
been extensively discussed in the literature
\citep{Yamamoto99,Suto99,Hamana01a,Hamana01b,Lahav04}, we simply
summarize the results below.

\subsubsection{Light-cone effect}

Since the current quasar sample extends from $z=0.72$ to $z=2.24$, the
clustering evolution within the survey volume cannot be neglected; for
instance, the linear growth rate of density fluctuations, $D(z)$, in the
universe with $\Omega_{\rm m}=0.27$ and $\Omega_\Lambda=0.73$ varies
from $D(0)/D(0.72)=1.41$ to $D(0)/D(2.24)=2.49$. This implies that
the corresponding matter 2PCF in the linear regime evolves by a factor  
of
$[D(0.72)/D(2.24)]^2 \approx 3.1$ between the two edges of the
survey volume. Although we could reduce the effect in principle by
dividing the entire sample into several subsamples in a narrow redshift
range, it would appreciably degrade the statistical significance of the
result. This is why one has to carry out the proper averaging procedure
in predicting $\xi_{\rm QSO}(s)$ theoretically. To be specific,
we follow the formula described by \citet{Hamana01a}:
\begin{equation}
\label{eq:xilc_qso1}
\xi_{\rm QSO}(s)
=\frac{\displaystyle{\int^{z_{\rm max}}_{z_{\rm min}}dz
\frac{dV_c}{dz}
\left[\phi(z) n_0(z)\right]^2\xi_{\rm QSO}(s,z)}}
{\displaystyle \int^{z_{\rm max}}_{z_{\rm min}}dz
\frac{dV_c}{dz}\left[\phi(z) n_0(z)\right]^2},
\end{equation}
where $z_{\rm min}$ and $z_{\rm max}$ denote the redshift range of the
survey, $dV_c/dz$ is the differential comoving volume element, $\phi(z)$
is the selection function of quasars, $n_0(z)$ is the number density of
quasars, and $\xi_{\rm QSO}(s,z)$ is the 2PCF of quasars in redshift
space at $z$.

In practice, we rewrite equation (\ref{eq:xilc_qso1}) as
\begin{equation}
\label{eq:xilc_qso2}
  \xi_{\rm QSO}(s)=
\frac{\displaystyle{\int^{z_{\rm max}}_{z_{\rm min}}dz
\left(\frac{dV_c}{dz}\right)^{-1}
\left(\frac{dN_{\rm QSO}(z)}{dz}\right)^2\xi_{\rm QSO}(s,z)}}
{\displaystyle \int^{z_{\rm max}}_{z_{\rm min}}dz
\left(\frac{dV_c}{dz}\right)^{-1}
\left(\frac{dN_{\rm QSO}(z)}{dz}\right)^2}.
\end{equation}
In the above expression, $dN_{\rm QSO}(z) / dz$ is the redshift number
distribution of quasars which is directly estimated from the data (Lower
panel of Fig.\ref{fig:M_z}), and $dV_c/dz$ can be easily computed once
a set of cosmological parameters is specified.

\subsubsection{Redshift-space distortion}

The next task is to convert the theoretical predictions in real space to
those in redshift space. Since the current paper focuses on scales
beyond $\sim 10h^{-1}$Mpc, the nonlinear effect (i.e., fingers of god)  
is
not important and we may just consider the Kaiser effect \citep{Kaiser}:
\begin{equation}
\label{eq:xis_qso}
\xi_{\rm QSO}(s,z)=\left(1+{2\over 3}\beta(z) + {1 \over 5} \beta^2(z)  
\right)
\xi_{\rm QSO}(r,z),\\
\beta(z) \equiv {1\over b_{\rm QSO}(z)}{d\ln D(z) \over d\ln a },
\end{equation}
where $r$ and $s$ denote the quasar pair-separations (comoving) in real
and redshift spaces, respectively, $a$ is the scale factor, and $b_{\rm
QSO}(z)$ is the biasing factor of quasars at $z$ as described below.
This expression assumes the distant-observer approximation, and can be
justified for pair-separations less than $10^\circ$ on the sky
\citep{matsubara00}. For reference, at the median redshift of $z\sim
1.4$, the comoving scale subtended by an angle of $10^\circ$ is $\approx
511 h^{-1}$Mpc in the $\Omega_{\rm m}=0.3$ and $\Omega_\Lambda=0.7$
model.  This is why we only use the bins less than 502 $h^{-1}$Mpc where
the distant-observer approximation can be reasonably justified.

\subsubsection{Quasar biasing}

The quasar biasing could be fairly complex in principle, but again in
the linear regime that we consider here, one can safely assume the
scale-independent and linear bias (e.g., \cite{matsubara99}):
\begin{equation}
\label{eq:xibias_qso}
  \xi_{\rm QSO}(r,z)=
\left[{b_{\rm QSO}(z)}\right]^2\xi_{\rm m}(r,z)
=\left[{b_{\rm QSO}(z)}\right]^2 D(z)^2 \xi_{\rm m}(r,0) .
\end{equation}

Then it should be emphasized here that all the above three effects (the
light cone effect, the redshift-space distortion, and the quasar
biasing) are linear in a sense that equations (\ref{eq:xilc_qso2}) to
(\ref{eq:xibias_qso}) yield a linear relation between the quasar 2PCF in
redshift space on the light-cone and the matter 2PCF in real space at
$z=0$.  In particular the three effects simply change the amplitude of
the matter 2PCF but none changes the positions of the baryon bump and
the zero-crossing scale.  This is valid only in the linear regime, and
significantly simplifies the analysis. Consequently we can simply
introduce an effective biasing factor, $b_{\rm eff}$:
\begin{equation}
\xi_{\rm QSO}(s) = {b_{\rm eff}}^2\xi_{\rm m}(r,0),
\end{equation}
and
\begin{equation}
  {b_{\rm eff}}^2 =
\frac{\displaystyle{\int^{z_{\rm max}}_{z_{\rm min}}dz
\left(\frac{dV_c}{dz}\right)^{-1}
\left(\frac{dN_{\rm QSO}(z)}{dz}\right)^2\left(1+{2\over 3}\beta(z) +  
{1 \over 5}
\beta^2(z) \right)b^2_{\rm QSO}(z)D^2(z)}}
{\displaystyle \int^{z_{\rm max}}_{z_{\rm min}}dz
\left(\frac{dV_c}{dz}\right)^{-1}
\left(\frac{dN_{\rm QSO}(z)}{dz}\right)^2}.
\end{equation}

We note a possible systematic effect that gives rise to scale-dependence
in the effective bias: if mean redshifts of quasar pairs which
contribute to small-scale and large-scale correlations are significantly
different, the biasing measured on the light-cone may acquire artificial
scale-dependence. In other words, the redshift dependence of the
selection function combined with that of the {\it scale-independent}
bias may induce additional scale-dependence in the effective bias.  For
instance, \citet{Tegmark04} discussed in detail how the luminosity
dependence yields such scale-dependence of biasing using a subset of the DR2
SDSS galaxy catalog.  In the current sample, however, this is not a
serious problem for two reasons.  First, while the luminosity and color
dependence of galaxy clustering is well established
\citep{Kayo-3pt,Zehavi04}, it is not yet clear that quasars also exhibit
similar strong dependence on their intrinsic properties.
\citet{2QZ-14}, for instance, find no significant luminosity dependence
in the clustering of 2dF QSOs. Second, as is clear from the QSO
distribution plot and $dN/dz$ (Figs.  \ref{fig:quasar-distribution} and
\ref{fig:wedge_quasar}), all separation bins of the 2PCF that we use in the
analysis are not dominated by those in any specific redshift range,
rather they are contributed equally over the entire redshift range. To be
more specific, we compute the mean redshifts of pairs, $\bar z$, at
separation bins of $s$ assuming $\Omega_{\rm m}=0.3$ and
$\Omega_\Lambda=0.7$. We find ($\bar z$, $s$) = (1.40, $13h^{-1}$Mpc),
(1.41, $25h^{-1}$Mpc), (1.41, $100h^{-1}$Mpc), (1.41, $200h^{-1}$Mpc),
(1.41, $400h^{-1}$Mpc), (1.42, $800h^{-1}$Mpc), and
(1.41,$1600h^{-1}$Mpc). This reasonably justifies the fact that we
neglect the luminosity dependence of quasar biasing.

Thus we proceed as follows; first we specify the values of $\Omega_{\rm
b}$, $\Omega_\Lambda$, and $h$, then compute $\xi_{\rm m}(r,0)$. Since
we assume the spatially flat model, $\Omega_{\rm m}$ is given as
$1-\Omega_\Lambda$. As discussed below, we do not have to specify the
value of $\sigma_8$, the top-hat mass fluctuation amplitude at
$8h^{-1}$Mpc, as long as we do not intend to solve for $b_{\rm QSO}(z)$.
Second we evaluate $\xi_{\rm QSO}(s)$ from the quasar sample using the
same values of $\Omega_{\rm m}$, $\Omega_\Lambda$, and $h$.  Finally, we
compute the $\chi^2$ of the observed and the model 2PCFs by varying the
value of $b_{\rm eff}$.  In doing so, we use both diagonal and
off-diagonal elements of the covariance matrix:
\begin{equation}
\chi^2 = \sum_{ij}
\left(\xi_{{\rm QSO,obs},i}- b^2_{\rm eff}\xi_{{\rm m},i}\right)
\left(\xi_{{\rm QSO,obs},j}- b^2_{\rm eff}\xi_{{\rm m},j}\right)
{C_{ij}}^{-1},
\end{equation}
where ${C_{ij}}^{-1}$ is the inverse matrix of the covariance matrix and
the subscripts, $i$ and $j$, are indices of separation bins.  We repeat
the above procedure for $0<\Omega_{\rm b}<1$ and $0<\Omega_\Lambda<1$,
and derive constraints on the two parameters. We adopt $h=0.7$ as a
fiducial value, but also show results for $h=0.6$ and $h=0.8$ as well,
although the $h$-dependence is very weak.

The resulting constraints are independent of the value of $\sigma_8$ and
also of the evolution model of $b_{\rm QSO}(z)$.  Nevertheless it is
also interesting to estimate the quasar biasing factor even in a
model-dependent manner. We adopt a simple evolution model by
\citet{Fry}:
\begin{equation}
\label{eq:frybias}
  b_{\rm QSO}(z) = 1+{b_{\rm QSO}(z=0)-1 \over D(z)},
\end{equation}
This model neglects the possible luminosity/color dependence of quasar
biasing and the finite lifetime of quasars, and may need to be improved
to some extent; we note that \citet{2QZ-14} find no significant
luminosity dependence in the clustering of 2dF QSOs.  Nevertheless
equation (\ref{eq:frybias}) gives us a useful measure of the expected
degree of the quasar spatial biasing at $z=0$.

\section{Results and Discussion}

In performing the statistical analysis, we recompute $\xi_{\rm
QSO,obs}(s)$ for different values of $\Omega_\Lambda$
($0<\Omega_\Lambda<1$ at an interval of $\Delta \Omega_\Lambda=0.05$).
Figure \ref{fig:sdss_lambda} shows the observed 2PCFs of the SDSS
quasars for several different values of $\Omega_\Lambda$ (the $n=1$
binning offset). Each data point and associated errorbar is slightly
shifted horizontally for clarity.  Unlike in Figures \ref{fig:baryon} to
\ref{fig:hubble}, the baryon bump is not directly visible due to the
large bin size. Because of the statistical limitation, we also do not
directly detect the anti-correlation that is predicted beyond a scale of
200$h^{-1}$Mpc (in the case of $\Omega_b=0.04$ and
$\Omega_\Lambda=0.7$ universe).  However, we find that there is
a break in the slope of the 2PCF around 100$h^{-1}$Mpc. While
$\xi_{\rm QSO,obs}(s)$ is always positive on the scales less than 100
$h^{-1}$Mpc, it is not inconsistent to be negative on the larger scale
for any sets of cosmological parameters that we surveyed here. Thus this
break may be interpreted to put a lower limit on the zero-crossing
scale of $\xi_{\rm QSO,obs}(s)$.

Figure \ref{fig:binning} shows the 2PCFs of SDSS quasars using the 6
different binning offsets ($\Omega_{\rm m} = 0.3$ and $\Omega_\Lambda =
0.7$).  The break of $\xi_{\rm QSO,obs}(s)$ around 100$h^{-1}$Mpc is
quite robust against the different choice of the binning.  Since the
data at different bins are strongly correlated, we do not use this
entire dataset for our analysis; this is just to show that the binning
makes little difference. In what follows, we perform statistical
analysis for the $n=1$ and $n=4$ offsets, separately.

Figure \ref{fig:chi2_bin030-1} shows the corresponding contours of
\begin{equation}
\label{eq:delta_chi2}
\Delta\chi^2(\Omega_{\rm b},\Omega_\Lambda)
\equiv \chi^2(\Omega_{\rm b},\Omega_\Lambda, b_{\rm eff, min})
-\chi^2_{\rm min},
\end{equation}
where $b_{\rm eff, min}$ is the value of $b_{\rm eff}$ which minimizes
$\chi^2$ for a given set of $\Omega_{\rm b}$ and $\Omega_\Lambda$, and
$\chi^2_{\rm min}$ is the global minimum value in $\Omega_{\rm b}$,
$\Omega_\Lambda$ and $b_{\rm eff}$ parameter space.  The three curves
represent the 68\%, 95\% and 99.7\% confidence levels.  Figure
\ref{fig:chi2_bin030-25} is the same plot but employs the $n=4$ binning
offset (see Fig. \ref{fig:binning}).

The cosmological parameters inferred from the WMAP results
\citep{Spergel03} are in agreement with our constraints in Figures
\ref{fig:chi2_bin030-1} and \ref{fig:chi2_bin030-25} at the 68\%
confidence level.  Given these uncertainties, therefore, we express our
constraints as allowed regions of $\Omega_{\rm b}$ and $\Omega_\Lambda$.
For this purpose, we fit the mean and $2\sigma$ contours as straight
lines and obtain
\begin{eqnarray}
\label{eq:n1fit_mean}
2.1\Omega_{\rm b} + \Omega_\Lambda = 0.89 \quad ({\rm mean}) , \\
\label{eq:n1fit_2sigma}
0.80- 2.8\Omega_{\rm b} < \Omega_\Lambda < 0.90 - 1.4\Omega_{\rm b}   
\quad (2\sigma)
\end{eqnarray}
for the $n=1$ binning offset (Fig. \ref{fig:chi2_bin030-1}), and
\begin{eqnarray}
\label{eq:n4fit_mean}
3.6\Omega_{\rm b} + \Omega_\Lambda = 0.99 \quad ({\rm mean}) , \\
\label{eq:n4fit_2sigma}
0.69 - 5.1\Omega_{\rm b} < \Omega_\Lambda < 1.0 - 2.3\Omega_{\rm b}  
\quad (2\sigma)
\end{eqnarray}
for the $n=4$ binning offset (Fig. \ref{fig:chi2_bin030-25}).  These
fitted lines are also plotted in Figures \ref{fig:chi2_bin030-1} and
\ref{fig:chi2_bin030-25}.  The differences of the slopes of contours
between the $n=1$ and $n=4$ binning offsets mainly come from the
smallest-scale behavior of $\xi_{\rm QSO, obs}(s)$, but the resulting
constraints are fairly insensitive to the different binning offsets
(compare Figs. \ref{fig:chi2_bin030-1} and \ref{fig:chi2_bin030-25}).
The dependence of our constraints on the Hubble constant is illustrated
in Figure \ref{fig:hubble_chi2}. The upper and lower panels are derived
by employing the $n=1$ and $n=4$ offsets, respectively.  The left and
right panels show the results for $h=0.6$, and $h=0.8$, and the central
panels plot the results for $h=0.7$ (identical to Figures
\ref{fig:chi2_bin030-1} and \ref{fig:chi2_bin030-25}) for comparison.
Except for the small difference, the overall conclusion is that our
constraints are very insensitive to the choice of $h$.

Finally, let us consider the corresponding biasing parameter of quasars
implied from the current study. Unfortunately the current sample cannot
constrain the redshift-dependence on the bias evolution.  Thus we have
to specify the value of $\sigma_8$ (at z=0) and the evolution model of
quasar biasing as well, although all the constraints presented above are
independent of those assumptions .  For simplicity and clarity, we adopt
$\Omega_{\rm b}=0.04$ and $\Omega_\Lambda=0.70$. Then we find that
$b_{\rm eff}$ is $1.13{\sigma_8}^{-1}$ for $0.6<\sigma<1.0$.  If we
define a ``mean'' quasar bias as
\begin{equation}
  {b_{\rm mean}}^2 = {b_{\rm eff}^2}\left[
\frac{\displaystyle{\int^{z_{\rm max}}_{z_{\rm min}}dz
\left(\frac{dV_c}{dz}\right)^{-1}
\left(\frac{dN_{\rm QSO}(z)}{dz}\right)^2\left(1+{2\over 3}\beta(z) +  
{1 \over 5}
\beta^2(z) \right)D^2(z)}}
{\displaystyle \int^{z_{\rm max}}_{z_{\rm min}}dz
\left(\frac{dV_c}{dz}\right)^{-1}
\left(\frac{dN_{\rm QSO}(z)}{dz}\right)^2}\right]^{-1},
\end{equation}
we find that $b_{\rm mean} = 1.59{\sigma_8}^{-1}$ (the mean redshift of the sample is $\bar z is 1.46$). This is a measure of
the degree of quasar clustering if the quasar biasing is {\it
independent of redshift}. We think this result is consistent with \cite{2QZ-14}.
If we adopt another bias evolution
(eq.[\ref{eq:frybias}]) instead, the current quasar sample is fitted to
have $b_{\rm QSO,Fry}(z=0)=0.54+0.83{\sigma_8}^{-1}$.

\section{Summary}
We have presented the first result of the SDSS quasar clustering
analysis. In particular we have focused on the behavior of the quasar
two-point correlation functions (2PCF) at large scales,
 and explored the embedded baryonic signature through the overall shape of the quasar 2PCF. 
The measurement is still statistically limited, but we found that
the zero-crossing scale of the quasar 2PCFs is larger than $100h^{-1}$Mpc.
The estimated quasar 2PCFs are in good agreement with those predicted in
linear theory combining the light-cone effect, the redshift-space
distortion and the scale-independent linear biasing model.

To proceed further, we have attempted to derive joint constraints on the
baryon density parameter $\Omega_{\rm b}$ and the cosmological constant
$\Omega_\Lambda$ under the assumption of the spatially flat universe.
Our constraints are approximately expressed by the relation: $0.80-
2.8\Omega_{\rm b} < \Omega_\Lambda < 0.90 - 1.4\Omega_{\rm b}$ at the
2$\sigma$ confidence level ($n=1$ binning offset).  While the statistical
significance is not competitive, the constraint is consistent with the
independent results from other observations such as WMAP
\citep{Spergel03} within the $1\sigma$ level.

The current analysis can be improved in many different ways; to consider
non-flat ($\Omega_{\rm m}+\Omega_\Lambda \not= 1$) models, to explore
a more general equation of state of the universe than the cosmological
constant (e.g., Yamamoto 2004), to consider the anisotropy of the
2-dimensional correlation function \citep{MS96}, to incorporate the
gravitational nonlinearity effect which is not important on $\sim
10h^{-1}$Mpc scales but still cannot be entirely neglected (Changbom
Park, private communication), to take into account the finite angular
size of the pair-separation (i.e., without adopting the distant-observer
approximation; see Matsubara 2000), and to elaborate on the statistical
modeling, in particular, the error estimation. We plan to perform such
improved analyses in due course.

\begin{figure}[p]
   \centering \FigureFile(140mm,140mm){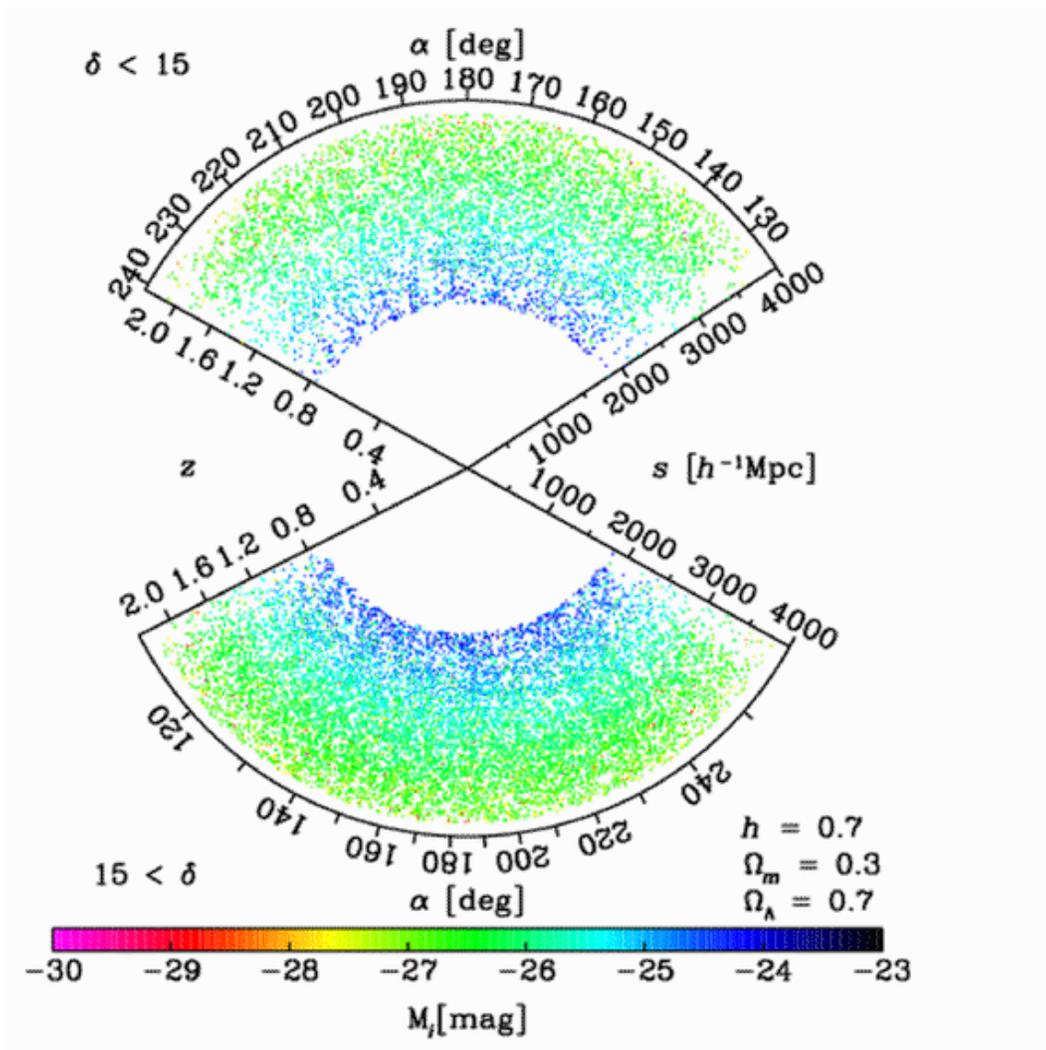}
\caption{Wedge diagram of the quasar distribution
(19986 quasars in total for $0.72<z<2.24$) corresponding to Figure
\ref{fig:quasar-distribution}.
\label{fig:wedge_quasar}}
\end{figure}

\begin{figure}[p]
   \centering \FigureFile(140mm,140mm){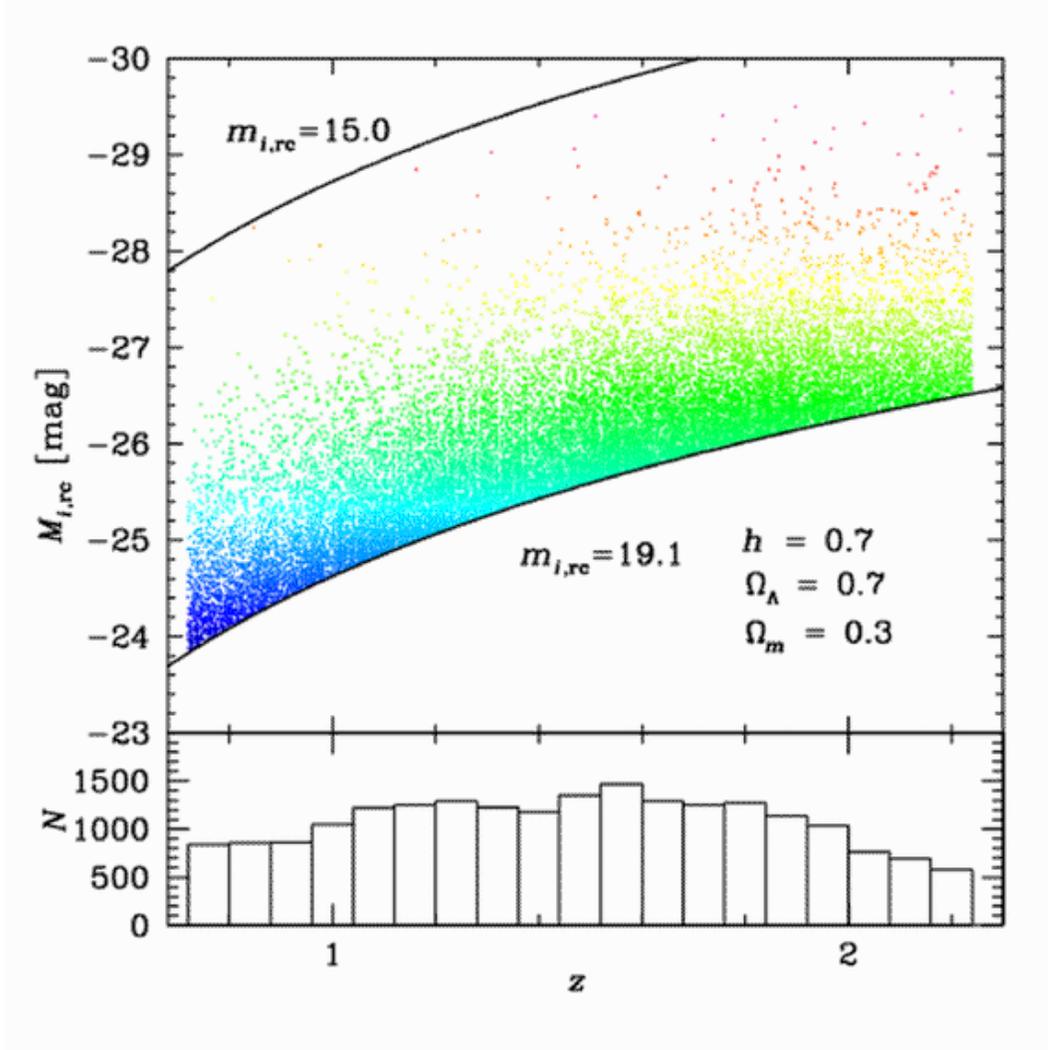}
\caption{ Scatter plot of redshift and $i$-band absolute magnitude ({\it
Upper panel}) and the corresponding redshift histogram ({\it Lower
panel}) of our quasar sample ($0.16 \le z \le 2.24$ and $15.0\le
m_{i,{\rm rc}} \le 19.1$). The width of the redshift bin is 0.08.
To compute the absolute magnitude, we assume
that the energy spectrum of quasars follows a single power-law with a
spectral index of $-0.5$, and adopt the cosmology of $\Omega_{\rm  
m}=0.3$ and
$\Omega_\Lambda=0.7$.
\label{fig:M_z}}
\end{figure}

The wedge diagram of the current quasar sample, Figure
\ref{fig:wedge_quasar}, clearly illustrates the isotropy and homogeneity
of the universe viewed over large scales, in marked contrast to the
complex clustering pattern of the local universe traced by galaxies
(e.g., Fig. 3 in \cite{Hikage03}). Therefore the most impressive lesson
that we have learned from the current analysis may be the fact that we
can indeed decipher cosmological information on $\Omega_{\rm b}$ and
$\Omega_\Lambda$ which is imprinted in the apparently almost homogeneous
distribution of quasars at such high redshifts.

\bigskip
We thank Masamune Oguri, Changbom Park and Edwin L. Turner for useful
discussions.  This research was supported in part by Grants-in-Aid for
Scientific Research from the Japan Society for Promotion of Science
(Nos.14102004 and 16340053). AJC is partially supported by an NSF CAREER
grant AST9984924 and an NSF ITR grant 1120201.

Funding for the creation and distribution of the SDSS Archive has been
provided by the Alfred P. Sloan Foundation, the Participating
Institutions, the National Aeronautics and Space Administration, the
National Science Foundation, the U.S. Department of Energy, the Japanese
Monbukagakusho, and the Max Planck Society. The SDSS Web site is
http://www.sdss.org/.

The SDSS is managed by the Astrophysical Research Consortium (ARC) for
the Participating Institutions. The Participating Institutions are The
University of Chicago, Fermilab, the Institute for Advanced Study, the
Japan Participation Group, The Johns Hopkins University, the Korean
Scientist Group, Los Alamos National Laboratory, the
Max-Planck-Institute for Astronomy (MPIA), the Max-Planck-Institute for
Astrophysics (MPA), New Mexico State University, University of
Pittsburgh, Princeton University, the United States Naval Observatory,
and the University of Washington.

\begin{figure}[p]
   \centering \FigureFile(140mm,140mm){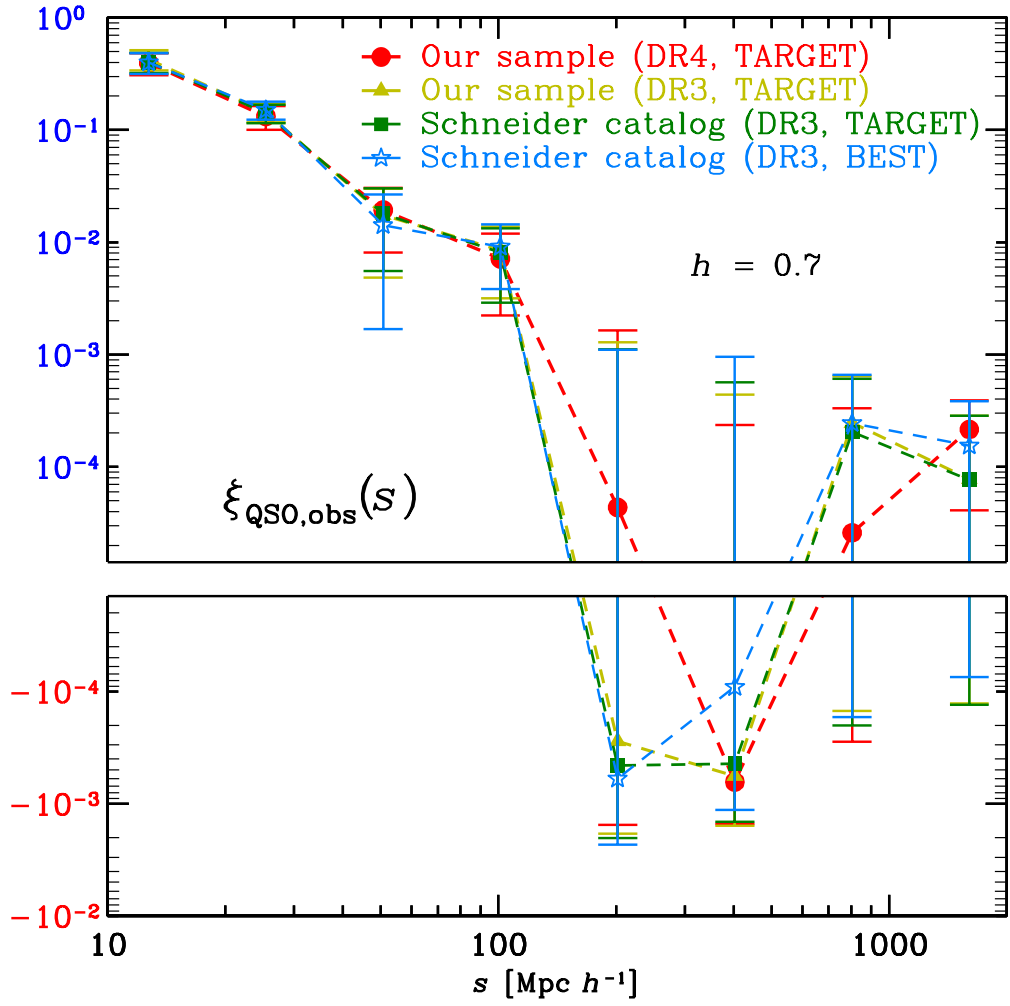}
\caption{The two-point correlation functions of our samples, BEST DR3
and TARGET DR3 (the $n=1$ binning offset) for $\Omega_{\rm m}=0.3$,
$\Omega_\Lambda=0.7$ and $h=0.7$.  Filled circles and triangles
correspond to our samples (DR4 and DR3), and filled squares and stars
correspond to the BEST and TARGET DR3 quasar samples in the catalog by
\citet{Schneider05}.  \label{fig:sdss_n1}}
\end{figure}
\begin{figure}[p]
   \centering \FigureFile(140mm,140mm){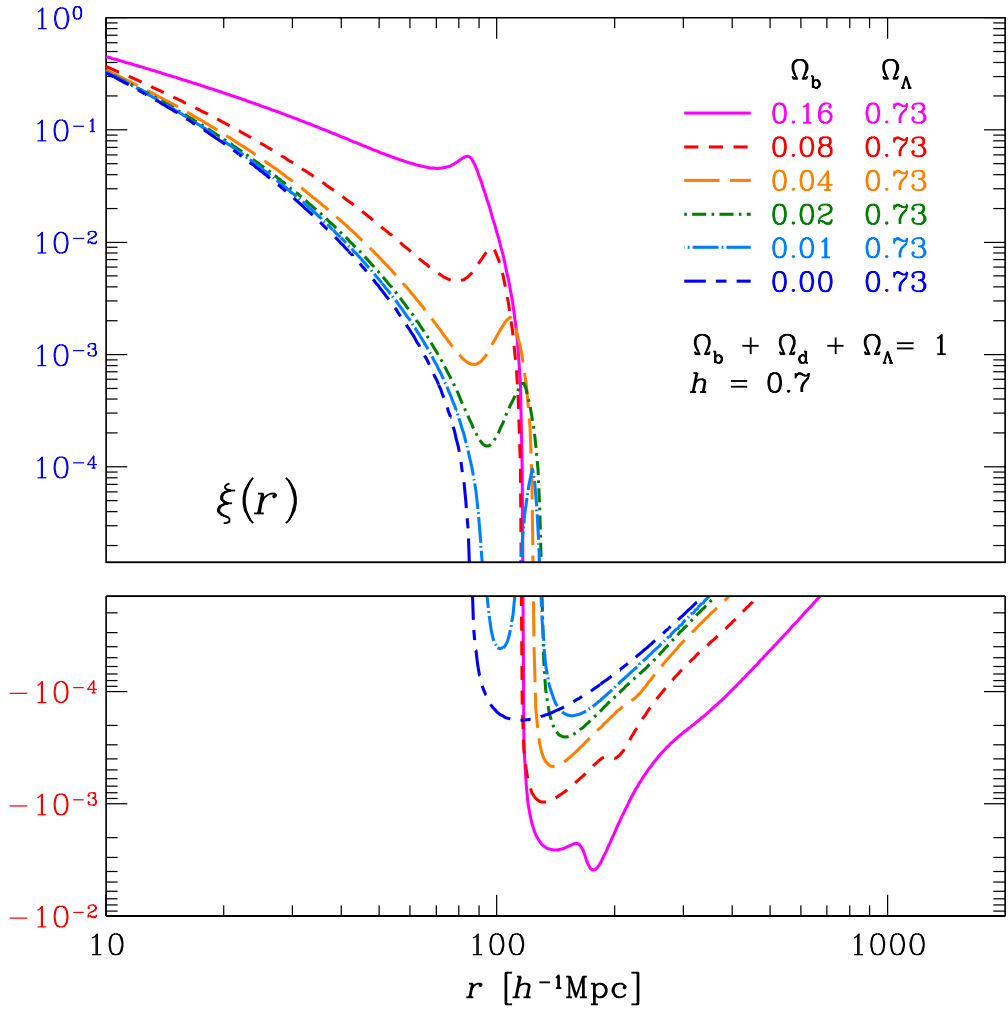}
\caption{Matter two-point correlation functions in linear theory (in
real space) at $z=0$ for different values of $\Omega_{\rm b}$.  The CDM
transfer function of \citet{EisensteinHu} is adopted.  We assume the
spatially flat model ($\Omega_{\rm b}+\Omega_{\rm  
d}+\Omega_\Lambda=1$), and
$h=0.7$.  \label{fig:baryon}}
\end{figure}
\begin{figure}[p]
   \centering \FigureFile(140mm,140mm){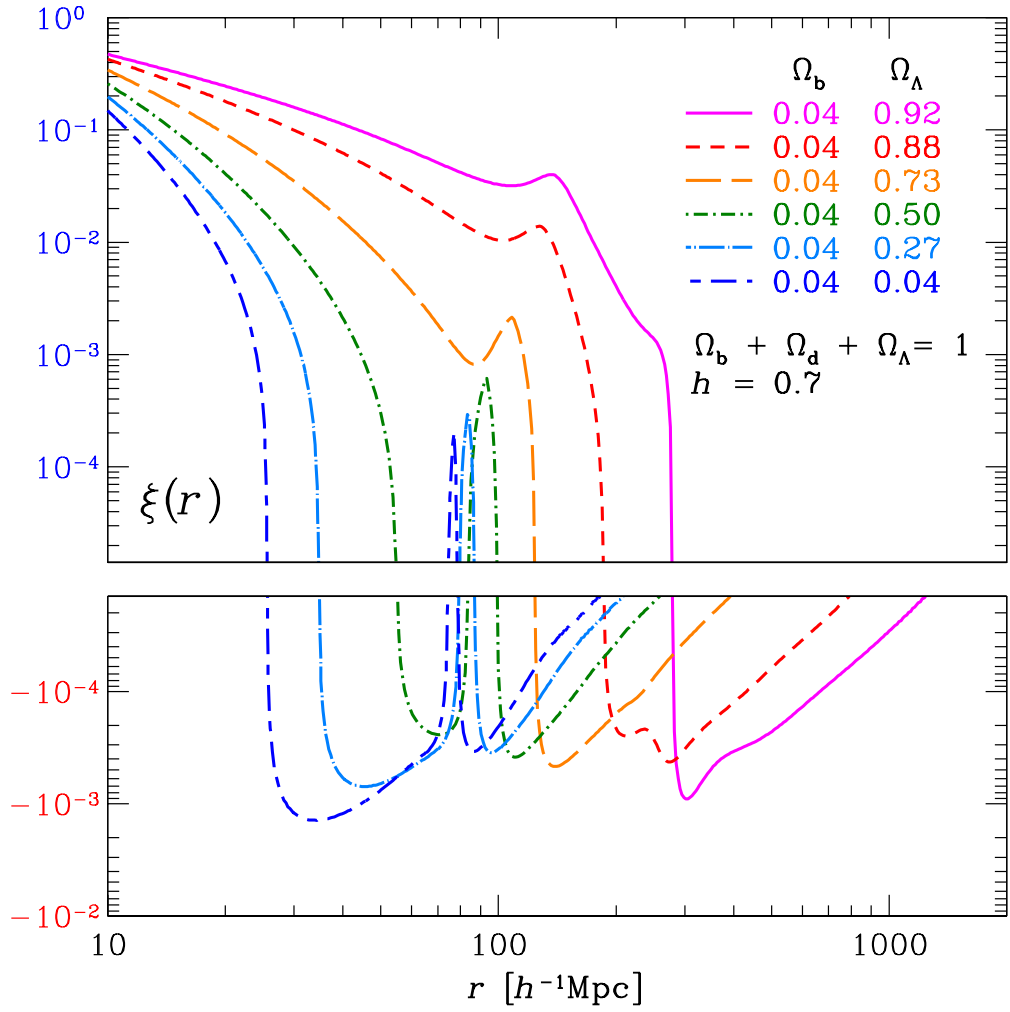}
\caption{Same as Figure \ref{fig:baryon} but for different values of
$\Omega_\Lambda$ with $\Omega_{\rm b}=0.04$ and $h=0.7$.
\label{fig:lambda}}
\end{figure}
\begin{figure}[p]
   \centering \FigureFile(140mm,140mm){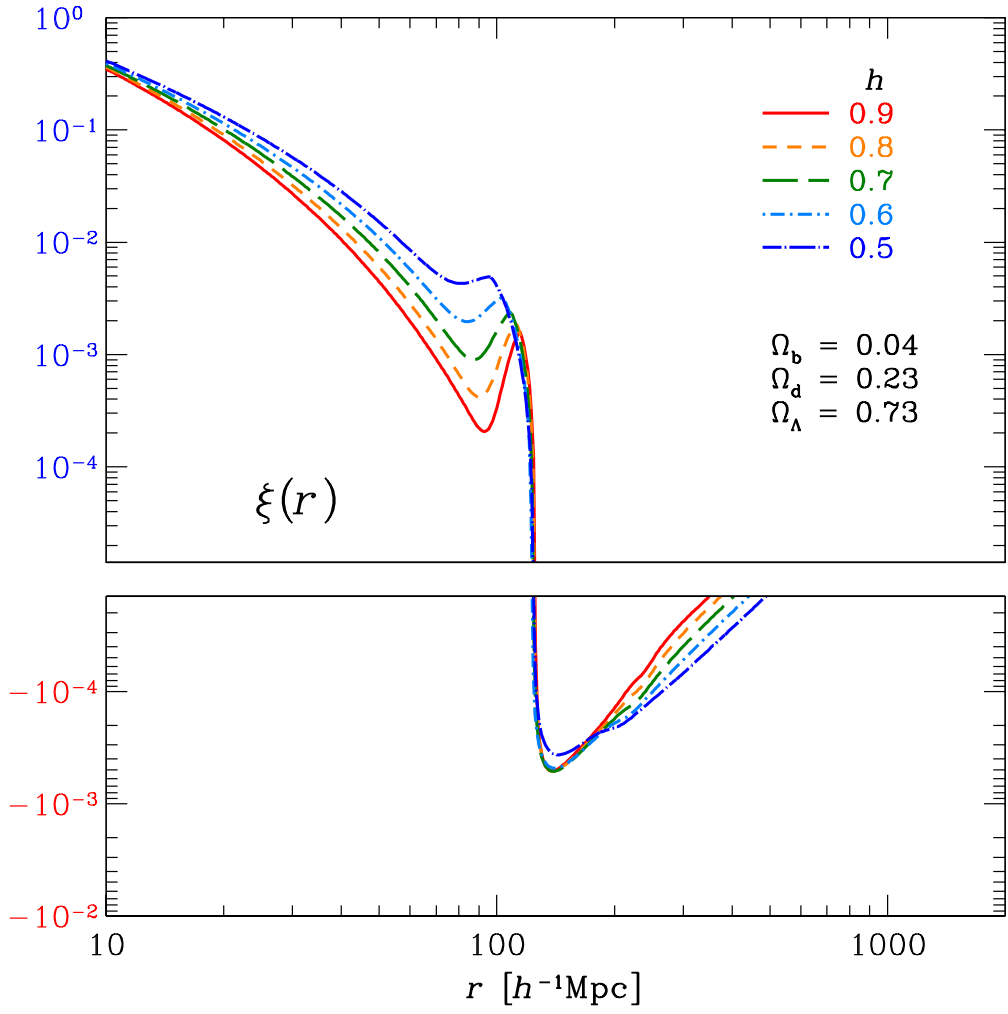}
\caption{Same as Figures \ref{fig:baryon} and \ref{fig:lambda} but for
different values of $h$ with $\Omega_{\rm b}=0.04$, $\Omega_{\rm  
d}=0.23$, and
$\Omega_\Lambda=0.73$.
\label{fig:hubble}}
\end{figure}
\begin{figure}[p]
   \centering \FigureFile(140mm,140mm){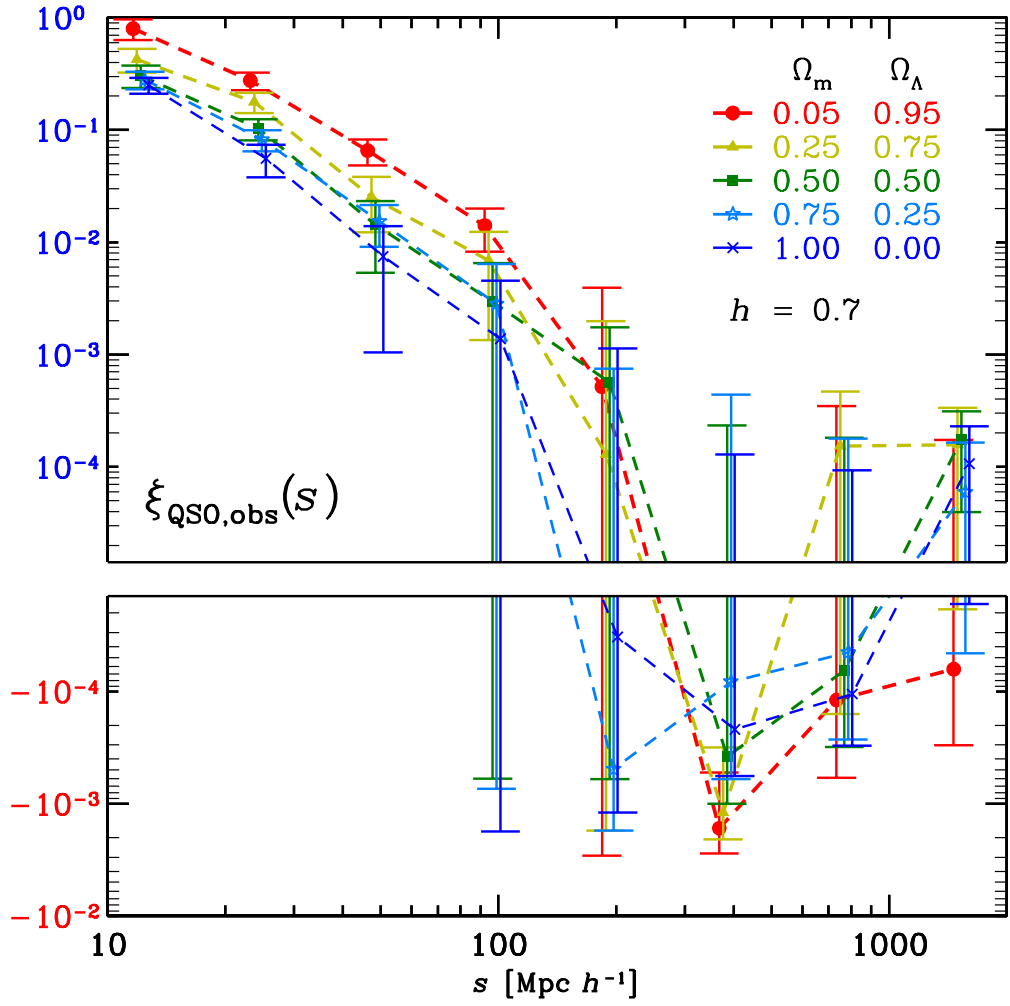}
\caption{The two-point correlation functions of SDSS quasars
(the $n=1$ binning offset) for several different
values of $\Omega_\Lambda$ ($h=0.7$).
  \label{fig:sdss_lambda}}
\end{figure}
\begin{figure}[p]
   \centering \FigureFile(140mm,140mm){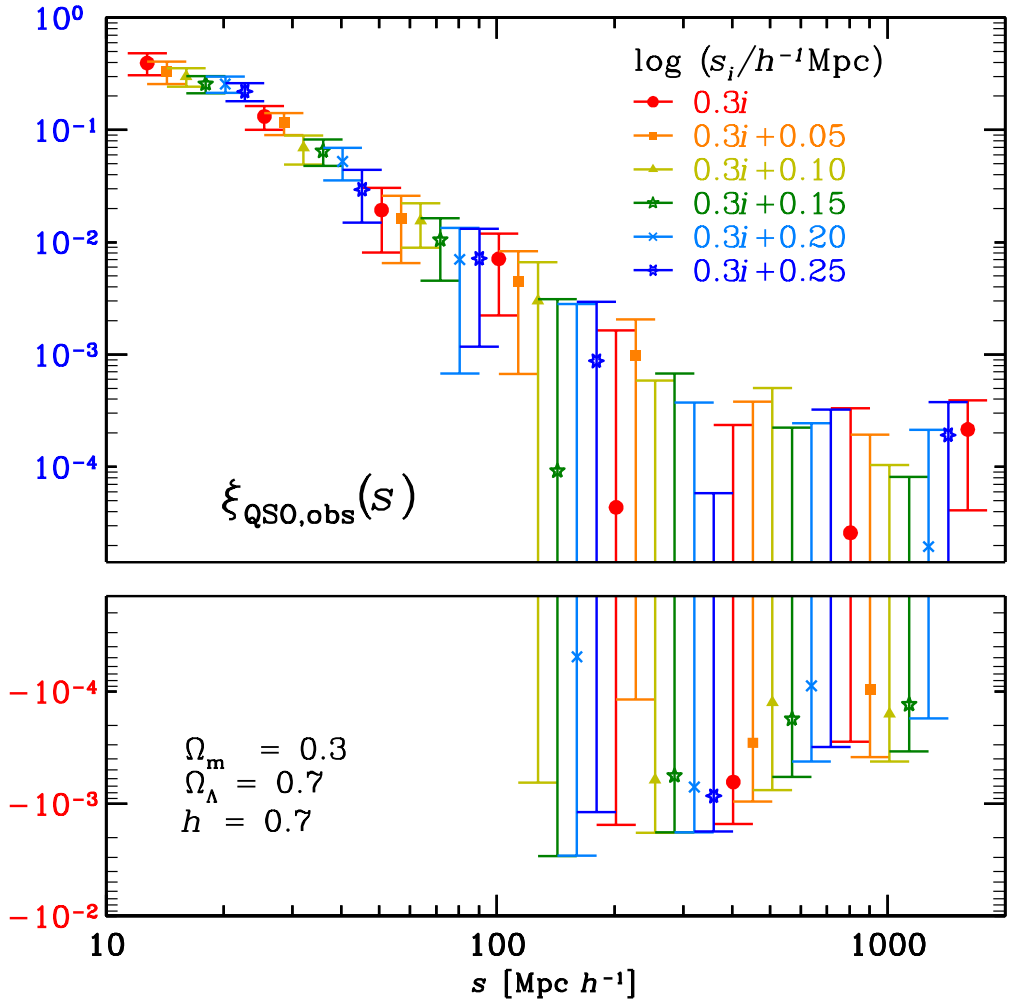}
\caption{The binning offset dependence of the two-point
correlation functions of quasars.
Results for the six different binning offsets (see eq.[\ref{eq:s_in}])
are plotted.  We adopt $\Omega_{\rm m} =0.3$,
$\Omega_\Lambda=0.7$, and $h=0.7$.
  \label{fig:binning}}
\end{figure}
\begin{figure}[p]
   \centering \FigureFile(140mm,140mm){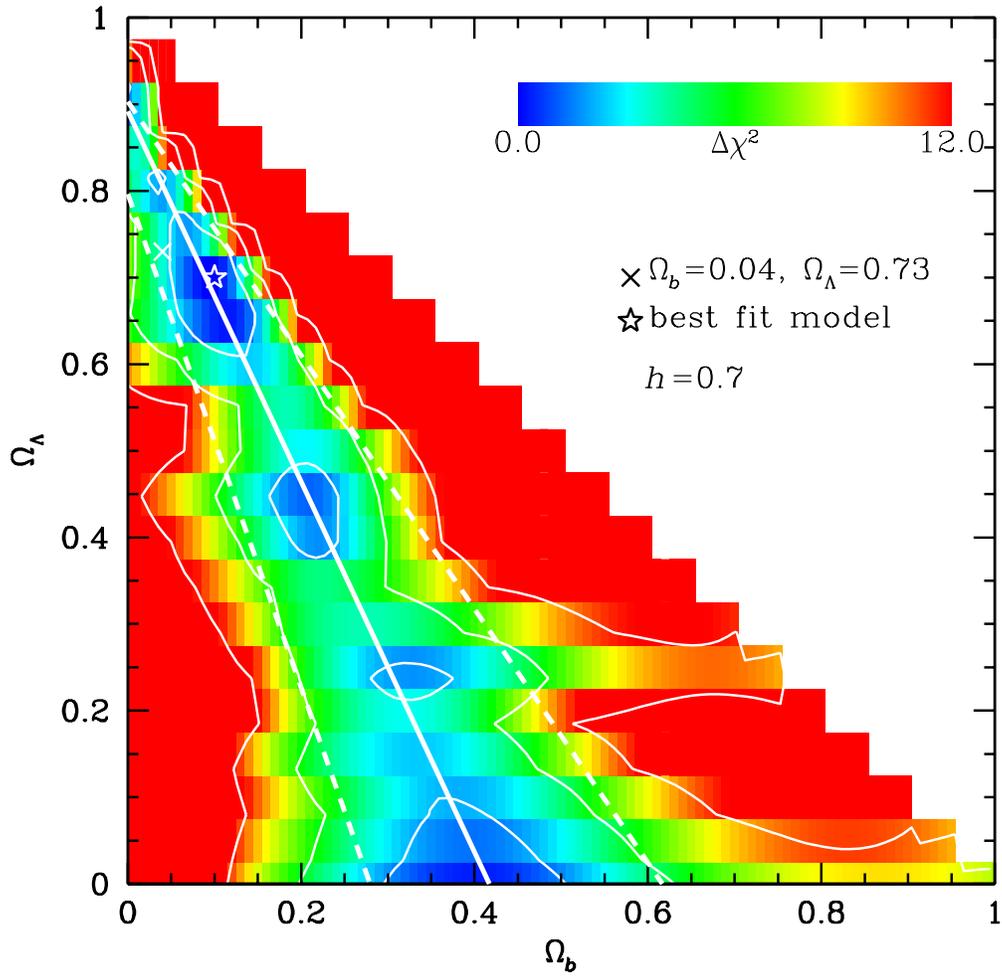}
\caption{Contour plot of $\Delta \chi^2$ (eq.[\ref{eq:delta_chi2}]) on
the $\Omega_{\rm b}$ -- $\Omega_\Lambda$ plane ($h=0.7$). The three  
contour
curves represent the 68\%, 95\% and 99.7\% confidence levels.  The cross
and the star indicate the values preferred by WMAP ($\Omega_{\rm  
b}=0.04$ and
$\Omega_\Lambda=0.73$), and our best-fit model ($\Omega_{\rm b} = 0.1$
and $\Omega_\Lambda = 0.7$). We adopt the $n=1$ binning offset in
equation (\ref{eq:s_in}).  The solid and dotted straight lines indicate
the linear fits to the mean and the $2\sigma$ contour levels
(eqs.[\ref{eq:n1fit_mean}] and [\ref{eq:n1fit_2sigma}]), respectively.
\label{fig:chi2_bin030-1}}
\end{figure}
\begin{figure}[p]
   \centering \FigureFile(140mm,140mm){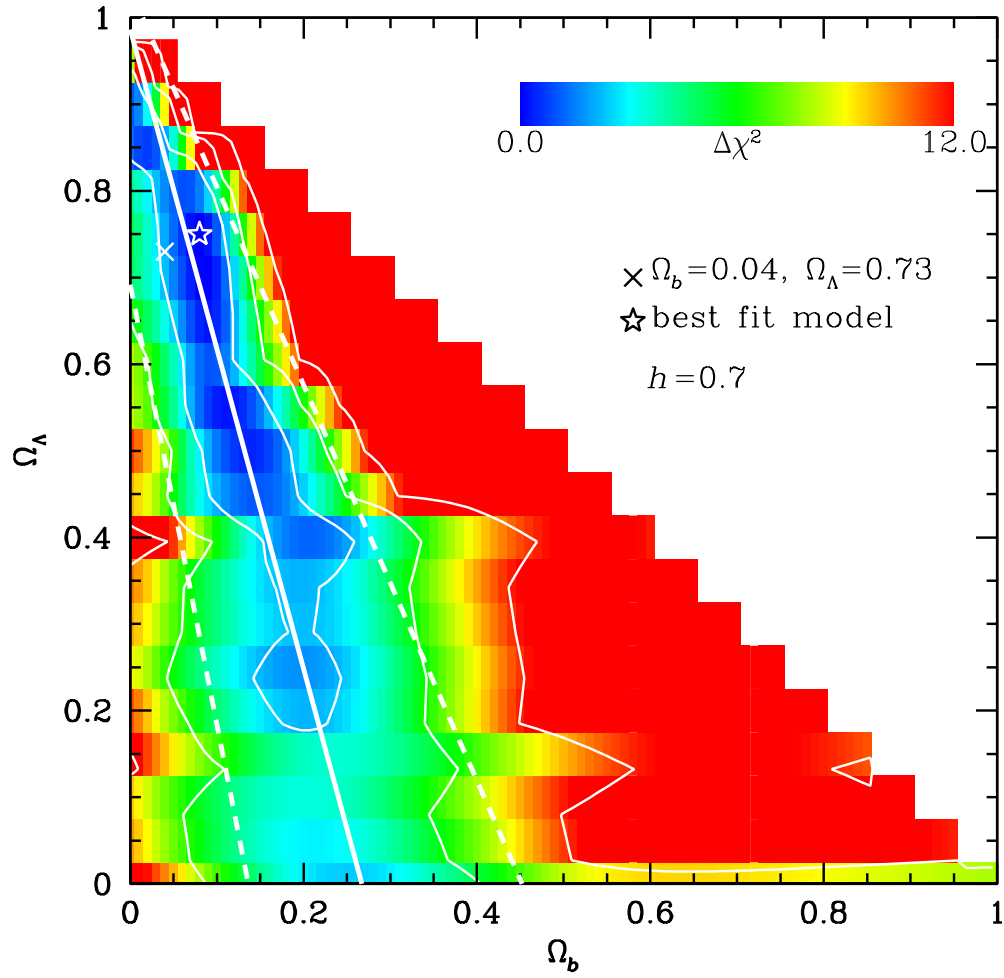}
\caption{Same as Figure \ref{fig:chi2_bin030-1} but for the $n=4$
binning offset.  The solid and dotted straight lines indicate the linear
fits to the mean and the $2\sigma$ contour levels
(eqs.[\ref{eq:n4fit_mean}] and [\ref{eq:n4fit_2sigma}]), respectively.
\label{fig:chi2_bin030-25}}
\end{figure}
\begin{figure}[p]
   \centering\FigureFile(140mm,102mm){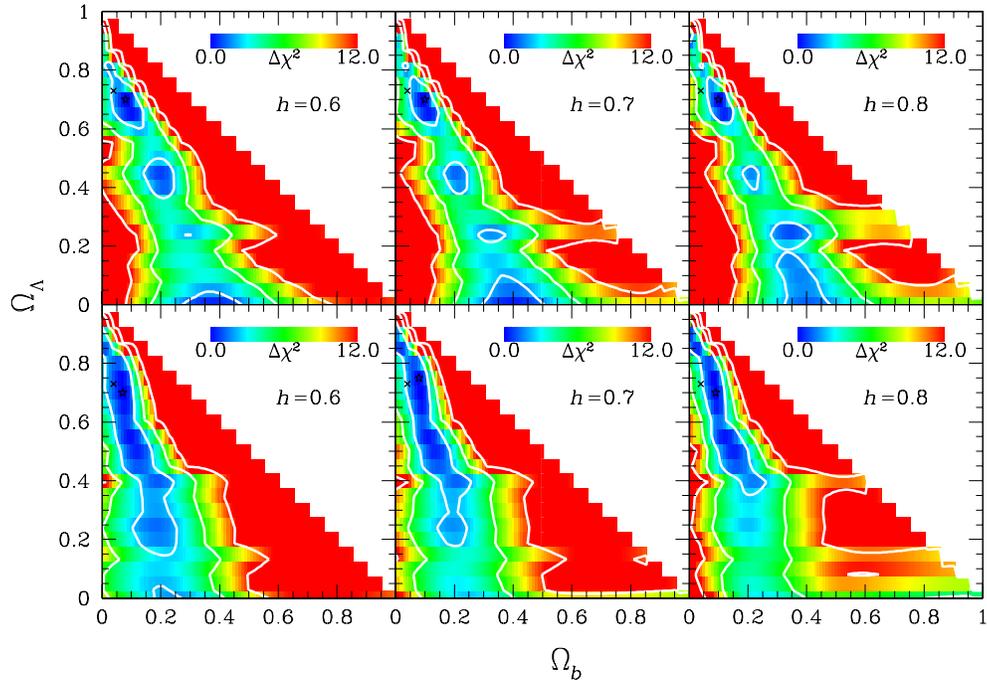}
\caption{Same as Figures \ref{fig:chi2_bin030-1} and
\ref{fig:chi2_bin030-25} but for $h=0.6$ ({\it Left panels}),
$h=0.7$ ({\it Center panels}),
and $h=0.8$ ({\it Right panels}).
  {\it Upper panels:} $n=1$ and
{\it Lower panels:} $n=4$ binning offsets.
\label{fig:hubble_chi2}}
\end{figure}

\bigskip

\begin{thebibliography}{}

\bibitem[Abazajian et al. (2004)]{dr3} Abazajian,~K., \etal\  \ 2004,
\aj, submitted (astro-ph/0410239)

\bibitem[Blanton et al.(2003)]{tiling} Blanton,~M.~R., Lupton,~R.H.,
Maley,~F.M., Young,~N., Zehavi,~I., \& Loveday,~J.  \ 2003, \aj, 125,
2276

\bibitem[Boyle, Mo (1993)]{Boyle1993} Boyle,~B.~J. \& Mo,~H.~J.  \
1993, \mnras, 260, 925

\bibitem[Connolly et al.(2005)]{connolly05}
Connolly,~A. \etal\ 2005, in preparation

\bibitem[Croom, Shanks (1996)]{Croom1996} Croom,~S.~M., Shanks,~T.  \
1996, \mnras, 281, 893

\bibitem[Croom et al.(2001)]{2QZ-2} Croom,~S.~M., Shanks,~T.,
Boyle,~B.~J., Smith,~R.~J., Miller,~L., Loaring,~N.~S. \& Hoyle,~F.  \
2001, \mnras, 325, 483

\bibitem[Croom et al.(2002)]{2QZ-9} Croom,~S.~M., Boyle,~B.~J.,
Loaring,~N.~S., Miller,~L., Outram,~P.~J., Shanks,~T., \& Smith,~R.~J.
\ 2002, \mnras, 335, 459

\bibitem[Croom et al.(2004)]{2QZ-14} Croom,~S.~M., \etal\
\ 2005, \mnras, 356, 415

\bibitem[Eisenstein, Hu (1998)]{EisensteinHu}
Eisenstein,~D.~J., \& Hu,~W.\ 1998, \apj, 496, 605

\bibitem[Fukugita et al.(1996)]{Fukugita1996} Fukugita,~M.,
Ichikawa,~T., Gunn,~J.E., Doi,~M., Shimasaku,~K., \& Schneider,~D.P.  \
1996, \aj, 111, 1748

\bibitem[Fry (1996)]{Fry} Fry,~N.~J., \ 1996, \apj, 461, L65

\bibitem[Gunn et al.(1998)]{Gunn}
Gunn,~J.~E., \etal\ 1998, \aj, 116, 3040

\bibitem[Hamana, et al.(2001)]{Hamana01a}
Hamana,~T., Colombi,~S., \& Suto,~Y.\ 2001, \aap 367, 18

\bibitem[Hamana et al. (2001)]{Hamana01b}
Hamana,~T., Yoshida,~N., Suto,~Y., \& Evrard,~A.~E. 2001 \apjl, 561, 143

\bibitem[Hikage et al.(2003)]{Hikage03} Hikage,~C., \etal\ 2003, \pasj,  
55, 911

\bibitem[Hogg et al.\ (2001)]{Hogg2001}
Hogg,~D.~W., Schlegel,~D.~J., Finkbeiner,~D.~P., \& Gunn,~J.~E. 2001,  
\aj, 122, 2129

\bibitem[Ivezic et al.\ (2004)]{Ivezic2004}
Ivezic,~Z., \etal\ 2004, AN, 325, 583

\bibitem[Kaiser (1987)]{Kaiser}
Kaiser,~N. 1987, \mnras, 227, 1

\bibitem[Kayo et al. (2004)]{Kayo-3pt}
Kayo,~I., \etal\  2004, \pasj, 56, 415

\bibitem[Kerscher, Szapudi \& Szalay (2000)]{Kerscher00}
Kerscher,~M., Szapudi,~I., \& Szalay,~A.~S. 2000, \apj, 535, L13

\bibitem[Landy, Szalay(1993)]{estimator}
Landy,~S.~D. \& Szalay,~S. 1993, \apj, 412, 64

\bibitem[Lahav, Suto (2004)]{Lahav04}
Lahav,~O. \& Suto,~Y. 2004, Liv. Rev. Rela.  7, 8

\bibitem[Lupton(1993)]{Lupton1993} Lupton,~R.~H. 1993, Statistics in
Theory and Practice (Princeton: Princeton Univ. Press)

\bibitem[Matsubara (1999)]{matsubara99}
Matsubara,~T. 1999, \apj, 525, 543

\bibitem[Matsubara (2000)]{matsubara00}
Matsubara,~T. 2000, \apj, 535, 1

\bibitem[Matsubara (2004)]{matsubara04}
Matsubara,~T. 2004, \apj, in press (astro-ph/0408349)

\bibitem[Matsubara,  Suto (1996)]{MS96}
Matsubara,~T. \& Suto, Y. 1996, \apj, 470, L1

\bibitem[Osmer (1981)]{Osmer1981}
Osmer.~P.~S. 1981, \apj, 247, 762

\bibitem[Ouchi et al.(2004a)]{Ouchi04a}
Ouchi,~M. \etal\ 2004a, \apj, 611, 660

\bibitem[Ouchi et al.(2004b)]{Ouchi04b}
Ouchi,~M. \etal\ 2004b, \apj, 611, 685

\bibitem[Outram et al.(2003)]{2QZ-11}
Outram,~P.~J., Hoyle,~F., Shanks,~T., Croom,~S.~M., Boyle,~B.~J.,  
Miller,~L., Smith,~R.~J. \& Myers,~A.~D.
2003, \mnras, 342, 483

\bibitem[Pier et al. (2003)]{Pier2003}
Pier,~J.~R., Munn,~J.~A., Hindsley,~R.~B., Hennessy,~G.~S.,
Kent,~S.~M., Lupton,~R.~H., \& Ivezic,~Z.
2003, \aj, 125, 1559

\bibitem[Pope et al.(2004)]{Pope04}
Pope,~A.~C., \etal\ 2004, \apj, 607, 655

\bibitem[Richards et al.(2002)]{targetselection}
Richards,~G.~T., \etal\ 2002, \aj, 123, 2945

\bibitem[Schlegel, Finkbeiner, Davis (1998)]{reddening}
Schlegel~D.~J., Finkbeiner,~D., \& Davis,~M. 1998, \apj, 500, 525

\bibitem[Schneider et al. (2005)]{Schneider05}
Schneider,~D.~P., \etal\ 2005, AJ, in press (astro-ph/0503679)

\bibitem[Shanks et al. (1987)]{Shanks1987}
Shanks,~T., Fong,~R., Boyle,~B.~J., \& Peterson,~B.~A. 1987, \mnras,  
227, 739

\bibitem[Shimasaku et al.(2003)]{Shimasaku03}
Shimasaku,~K. \etal\ 2003, ApJ, 586, L111

\bibitem[Smith et al. (2002)]{Smith2002}
Smith,~J., \etal\  2002, \aj, 123, 2121

\bibitem[Spergel et al. (2003)]{Spergel03}
Spergel,~D.~N. \etal\ 2003, \apjs, 148, 175


\bibitem[Steidel et al.(1999)]{Steidel99}
Steidel,~C.~C., Adelberger, K.~L., Giavalisco,~M.,
Dickinson,~M., \& Pettini,~M., 1999, ApJ, 519, 1

\bibitem[Stoughton et al. (2002)]{EDR}
Stoughton,~C., \etal\ 2002, \aj, 123, 485

\bibitem[Suto et al.(1999)]{Suto99}
Suto,~Y., Magira,~H., Jing,~Y.~P., Matsubara,~T., \&
Yamamoto,~K., 1999, Prog. Theor. Phys. Suppl., 133, 183

\bibitem[Tegmark et al.(2004)]{Tegmark04}
Tegmark,~M., \etal\  2004, \apj, 606, 702

\bibitem[Yamamoto (2004)]{Yamamoto04}
Yamamoto,~K. 2004, \apj, 605, 620

\bibitem[Yamamoto, Suto (1999)]{Yamamoto99}
Yamamoto,~K., \& Suto,~Y., 1999, \apj, 517, 1

\bibitem[York et al. (2000)]{Y2000}
York,~D.~G.,  \etal\ 2000, \aj, 120, 1579

\bibitem[Zehavi et al. (2004)]{Zehavi04}
Zehavi,~I. \etal\ 2004, submitted to \apj (astro-ph/0408569)

\end{thebibliography}
\end{document}